\documentclass[12pt,preprint]{emulateapj}

\usepackage{natbib,amsfonts}
\newif\ifAMStwofonts
\AMStwofontstrue
\newcommand{\hr}{$^{\rm h}$}
\newcommand{\mn}{$^m$}
\newcommand{\secs}{$^s$}

\newcommand{\hi}{H{\sc i}}
\newcommand{\kms}{\mbox{km s$^{-1}$}}
\newcommand{\gesim}{\raisebox{-0.4ex}{$\stackrel{>}{\scriptstyle\sim}$}}

\newcommand{\caplesim}{$\stackrel{<}{\scriptstyle\sim}$}

\shortauthors{Muller et al.}
\received{2004 May 27}
\begin{document}
\title{A Statistical Investigation of \hi\ in the Magellanic Bridge}
\author{E. Muller} \affil{Arecibo Observatory, HC3 Box 53995, Arecibo,
  Puerto Rico 00612} \email{emuller@naic.edu}
\author{S. Stanimirovi\'c, E. Rosolowsky} \affil{Radio Astronomy Lab, UC
  Berkeley, 601 Campbell Hall, Berkeley, CA 94720} \and \author{L.
  Staveley-Smith} \affil{Australia Telescope National Facility, CSIRO, PO Box 76, Epping, N.S.W. 1710,
  Australia}

\begin{abstract}
  We present results from two statistical analyses applied to an
  neutral hydrogen (HI) dataset of the nearby tidal bridge 
in the Magellanic System.
  Primarily, analyses of the Spatial Power Spectrum suggest that the
  Magellanic Bridge, historically considered to be a single contiguous
  feature, may in fact be a projection of two kinematically and
  morphologically distinct structures. The southern and more obviously
  turbulent parts appear to show structure organized similarly to the
  adjacent Small Magellanic Cloud (SMC), while the northern regions are 
shown to
  be relatively deficient in large scale power.  The extent of modification
  to the spatial power index by the velocity fluctuations is also highly
  variant across these parts of the Bridge. We find again that the northern
  part appears distinct from the southern part and from the SMC, in that the
  power spectrum is significantly more affected by slower velocity
  perturbations.

  We also probe the rate of spectral variation of the \hi\ by measuring the
  Spectral Correlation Function over selected regions.  The results from
  this analysis highlight a tendency for the \hi\ spectra within the bright
  parts of the Bridge to have a more persistent correlation in the E-W
  direction than in the N-S direction. These results are considered to be
  quantitative evidence for the tidal processes which are thought to have
  been active throughout the evolution of the Magellanic Bridge.
\end{abstract}


\section{Introduction}
\label{sec:introduction}

The Magellanic Bridge (MB) is an elongated, predominantly neutral
hydrogen (\hi) structure, located between and joining the Large and
Small Magellanic Clouds (Kerr, Hindman, \& Robinson, 1954). Numerical
simulations (Ru\v zi\v cka, 2003; Gardiner, Sawa, \& Fujimoto,
1994; Gardiner \& Noguchi, 1996) and the relative geometrical
arrangement of the components of the Magellanic System
(e.g. Mathewson, Schwarz \& Murray, 1977) indicate that the MB was
formed directly from the Magellanic Clouds, following a particularly
close mutual pass $\sim$200 Myr ago.  Recent high resolution
observations have shown that the \hi\ in the MB has a complex and
chaotic filamentary structure (Muller et al. 2003). Apparently
coherent \hi\ features are identifiable across the entire range of
observed angular scales, from $\sim$98$\arcsec$\ up to $\sim$7$\degr$,
although the filamentary component is clearly unevenly distributed
throughout the entire MB region.

Statistical techniques are a useful means to compare populations of
similar objects in different systems, understand and model general
properties and behaviors, and expose and isolate any hidden underlying
trends within a dataset.  In particular, statistical analyses provide
some of the most useful descriptions of turbulent flows. The results of
statistical analyses of emission-line data sets can be
directly compared with theoretical predictions to determine the properties
of the turbulence under that theoretical framework.  While there are many
statistical techniques that have been used to analyze different 
data sets, this paper focuses on two complimentary methods: 
the Spatial Power Spectrum (SPS) and the Spectral Correlation Function
(Rosolowsky et al, 1999).

The SPS and the SCF were chosen because they highlight the importance of structure
across a wide range of spatial scales.  The organisation of spatial scales has a
direct influence on the efficiency of energy transport throughout the
interstellar medium (ISM) of the subject system and also vice versa: 
the transport of energy
is an important operator on the arrangement and distribution of
spatial scales.  

The relationship between the
observed SPS and the characteristics of the underlying turbulent flows
has been the work of several theoretical studies (Lazarian \& Pogosyan
2000; Goldman 2000; Miville-Deschenes et al. 2003).  In contrast
with the SPS which measures structure of data sets on a more global scale,
the SCF measures local variations as a function of spatial scale.  

No studies have yet made a direct and dedicated comparison of the
outcomes and predictions of the SPS and the SCF. Measurements of the SCF on the
\hi\ of the LMC (Padoan et
al. 2001) were shown to be in rough, though not particularly good
agreement with earlier studies using the SPS (Elmegreen et al. 2001)
and a more detailed discussion of the predictions of the SCF in the context
of the turbulent and small-scale structure of the LMC \hi\  were not entered into.

To date, studies of the SPS have been made of \hi\ in the SMC and the LMC
(Stanimirovic et al. 1999a; Stanimirovic \& Lazarian 2001; Elmegreen et
al. 2001). SCF studies have been made only of the LMC (Padoan et
al. 2001). These results have highlighted the existence of important active
processes and structure trends within these systems and have been used as a quantifier for
the organisation of structure within these systems independently. The MB
forms a physical link between the SMC and the LMC, and represents a feature
which has evolved purely in response to tidal forces. The MB is therefore an
ideal system to test and compare the response of the SCF to a) a system
which has developed in response to large-scale forces and b) in the context of the
host systems, i.e. the SMC and LMC. Such an analysis may characterize the turbulence
in tidal features in general and in addition, a comparison of the SPS results 
derived for different components in the Magellanic System can
provide insights into different evolutionary paths for organization of
\hi\ structures in an galactic versus an intergalactic medium.

The structure of this paper is organized as follows.  In Section 2 we
review the SPS and SCF. In Section~\ref{sec:dataset} we summarize the
observations of the subject \hi\ MB dataset.  Sections~\ref{sec:sps}
and~\ref{sec:calculating scf} detail the methodology 
and results of the calculation of the \hi\ SPS and SCF on the MB \hi\
dataset. Sections~\ref{sec:sps_disc} presents the discussion and
interpretation of our results.  Finally, we summarize our results in
Section~\ref{sec:discussion}.

\section{Statistical Techniques}
\label{techniques}

\subsection{The Spatial Power Spectrum}

The SPS has traditionally been applied to obtain a useful insight into
hydrodynamics, MHD theory and turbulent motions (e.g. Lazarian \& Pogosyan
2000). 
For example, a shallow SPS
results from a system which is dominated by small-scale structure, while a steep SPS
results from a system which is dominated by large-scale fluctuations.
The SPS has been commonly applied on large datasets in an attempt to
parameterise the structure hierarchy of the \hi\ in these systems. 
(Crovisier \& Dickey 1983; Green 1993;
Stanimirovi\'c et al. 1999a; Stanimirovi\'c \& Lazarian 2001;
Elmegreen, Kim, \& Staveley-Smith 2001; Dickey et al.  2001).
  
The SPS is derived from the Fourier transform power spectrum of an
observed 2-d image, denoted $P(k)$, where $k$ is the spatial
frequency. Different types of observations commonly show that the SPS 
of intensity fluctuations can be characterized by a power-law, or
the sum of two power laws:
\[P(k)\propto k^{\gamma} \mbox{ or } P(k)\propto C_1k^{\gamma_1}+C_2k^{\gamma_2}.\]
The composite power law is observed when the dynamics of a system are shaped by
a process (or processes) which generate excess power over a specific 
range of scales. For example, supernovae are capable of injecting power across scales
of tens of parsecs.  The constants $C_1$ and $C_2$ are dependent on the relative importance of the two terms (see also Elmegreen \& Scalo, 2004; Scalo \& Elmegreen,
2004, for a detailed review of turbulent processes).

In the case of a Kolmogorov-type of turbulent fluid (isotropic and
incompressible turbulence, characterized by energy injection on large
scales with a dissipationless cascade of energy  
over intermediate scales and dissipation of energy only on smallest
scales) the SPS of both 3-d density and velocity fluctuations is a power law
with a slope equal to $-11/3$.

However, a spectral index of $\gamma\sim -3.7$ does not necessarily imply
Kolmogorov-like turbulence.  Recent work using simulated datasets have
shown that other turbulent regimes are capable of producing spectral
indices which are approximately equal to this value.
For example, Miville-Desch\^enes, Levrier \& Falgarone (2003) have developed
synthetic fractal Brownian motion (fBm) datasets which can yield spatial 
spectral indices of $-$3.7. Furthermore, synthetic datasets containing
randomly distributed, supernovae-driven expanding shells also appear
to produce a Kolmogorov spectrum (Hodge \& Deshpande, in
preparation). Both of these studies are in need of further
clarification however, specifically in that the fBm set does not represent
coherent turbulent structures (i.e. it is constructed from Brownian noise), and that the Hodge \&
Deshpande analysis involves unphysical formation processes (i.e. they do not
solve the hydrodynamic equations in the development of the set).  To distinguish and disentangle the nature and extent of velocity fluctuations
from static hierarchical density structure, techniques which measure the
velocity statistics of the set must be used.

An important advantage the SPS has over other 
statistical techniques is in relating  statistics of the observed intensity fluctuations with 
the statistics of the underlying 3-d density and velocity fields. A recent development in SPS studies involves the averaging of data across a
number of observed velocity channels. By integrating a spectral-line data
cube over varying ranges of velocity, Lazarian \&
Pogosyan (2000) showed that it is possible to disentangle  the influences of density and velocity fluctuations in the turbulent
flow, i.e. the power-law index is a function of both the
velocity, and the velocity width of the observations: $\gamma=\gamma(v,\Delta v)$.  When an
image is generated from a small range in velocities (small $\Delta
v$), a significant fraction of the observed structure results from the
velocities of the turbulent flow moving emission into different
velocity channels.  Since an SPS is usually inferred to characterize
the density fluctuations in a turbulent flow, the velocity
fluctuations would corrupt the results.  However, when integrated
over a large velocity range (large $\Delta v$), most of the velocity
fluctuations average out, leaving only static density fluctuations.
By
measuring the power law spectrum over a range of widths in the
velocity window, Lazarian \& Pogosyan (2000) showed that it was
possible to separate the power-law spectrum of the density
fluctuations from those of the velocity fluctuations.  In their
Velocity Channel Analysis (VCA), they refer to the regime where the
velocity thickness of the integrated slice averages out the velocity
fluctuations as the ``thick slice regime'' and the opposite regime
where the velocity fluctuations are important as the ``thin slice
regime''.

\subsection{The Spectral Correlation Function}

As the SPS is derived from the Fourier transform of the data set,
information from the entire image contributes to the power
spectrum.  However, this global approach compromises the usefulness of the SPS in isolating local
features in the data.  

By contrast, the SCF does not transform the data and operates on entire \emph{spectra} in a data cube
instead of entire images.  The SCF parameterizes the `similarity',
$S_0$, between two spectra as a function of their spatial separation,
$\Delta r$ and often shows a power-law relationship $S_0\propto
\Delta r^{\alpha}$ over a range of separations (e.g. Padoan, Rosolowsky \& Goodman, 2001).  The SCF is normalized
so that a perfect match between two spectra will yield a value of
$S_0=1$, and a value of zero is obtained for no similarity, or for an
anti-correlation. Since this technique does not use a Fourier
transform it does not require the data to be edge tapered and thereby
reduces problems arising from the finite sampling and the associated
edge effects. Therefore, the minimum size of the region that may be
sampled with this technique is much smaller than that available to the
power spectrum analysis.  However, the SCF is sensitive to the noise
in the data cube and its application is limited to high signal-to-noise
data. 

The early forms of the SCF technique were used to derive an estimate
of spectral similarity over a lag range which was defined by the
velocity dispersion of the spectra of interest. It was therefore
limited to data cubes that were described by a simple, single velocity
component. Later modifications to the SCF made by Padoan, Rosolowsky
\& Goodman (2001) included a parameter which weights the function by
the ratio of the RMS of the signal and noise.  These modifications
improved the ability of the algorithm to handle noise and expanded its
application to more complex and multi-component spectra and across any
range of spatial lags (Padoan, Rosolowsky \& Goodman 2001).  An
interesting application of the SCF was shown in the study by Padoan et
al. (2001) where a break in the power-law behavior of the SCF was
interpreted as a signature of the LMC's disk thickness.

As the SCF is a rather recent development, the relationship between the SCF results and the
properties of the observed turbulence are not understood as completely as
for the SPS. We have attempted here a more dedicated study and
comparison of the the SCF in the context of predictions from the
relatively well-parameterized SPS.

\section{The MB \hi\ dataset}
\label{sec:dataset}

We use the high-resolution \hi\ observations of Muller et al.  (2003)
for the present statistical investigation of the \hi\ distribution in
the MB. These observations encompass a $\sim7\degr \times 7\degr$
region, centered on RA = 2\hr 08\mn 40\secs, Dec. =
$-$73$\degr$27$\arcmin$59$\arcsec$\ (J2000), and covering the
heliocentric velocity range 100--350 \kms.  The data were collected
with both the Australia Telescope Compact Array (ATCA) and the Parkes
Telescope\footnote{The Australia Telescope Compact Array and Parkes
telescopes are part of the Australia Telescope which is funded by the
Commonwealth of Australia for operation as a National Facility managed
by CSIRO.}.  The observing parameters, data reduction and data
combination procedures are described in Muller et al. (2003).  The
final \hi\ data cube has 1-$\sigma$ brightness temperature sensitivity
of $\sim$0.9 K (corresponding to a column density of
$\sim$1.62$\times$10$^{18}$ cm$^{-2}$), a velocity resolution of
$\sim$1.6 \kms, and an angular resolution of 98$\arcsec$.  These
observations are sensitive to all spatial scales from about 29 pc
(corresponding to angular scale of $\sim$98$\arcsec$) to 7.3 kpc
(corresponding to angular scale of $\sim$7$\degr$), assuming a
distance of 60 kpc to the MB. An important note in the context of this work,
is that the ATCA interferometer, being composed of discrete elements is not
capable of uniformly sampling the entire UV plane. The missing data is
extrapolated during the deconvolution and inversion process which also
homogonises the sensitivities somewhat (refer to Muller et al. 2003 and also
Stanimirovi\'c, 1999 for explicit information on the data deconvolution and
inversion process).

\begin{figure*}
\plotone{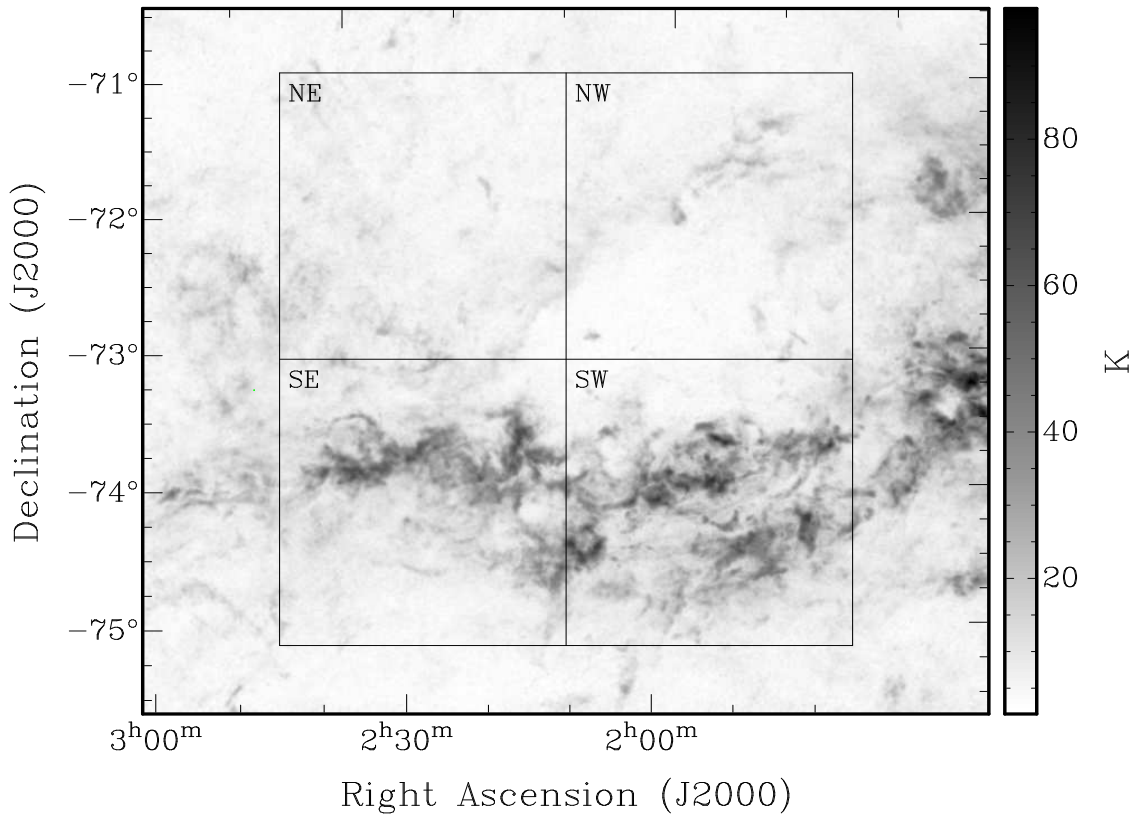}
\caption{\label{f:himap}Peak intensity map
  of the entire MB \hi\ dataset. The overlaid square indicates the four 256
  $\times$ 256 pixel ($\sim$2.2$\times$2.2 kpc) subsections over which the
  power spectrum analysis is calculated.}
\end{figure*}

\begin{figure*}
\epsscale{0.4}
\plotone{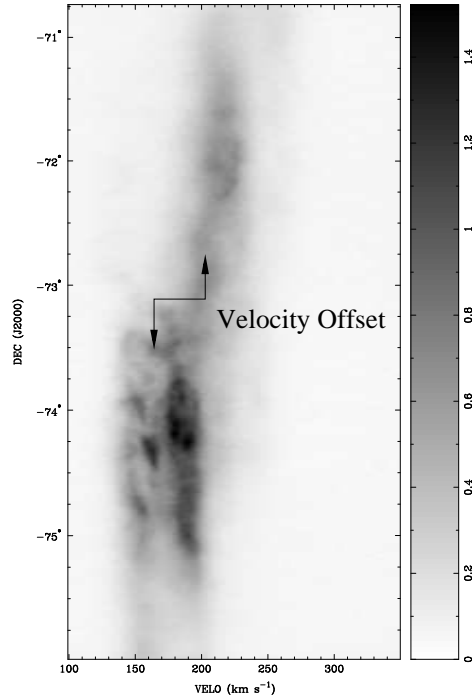}
\caption{\label{f:vel-dec}Integrated intensity map of the
Declination-Velocity projection of the combined ATCA-Parkes data cube.
The northern part of the \hi\ in the Bridge is clearly offset to
higher velocities with respect to the southern part by $\sim$40-50
\kms. We see a clear bimodal distribution of \hi\ in the more southern part
in this figure. This apparent two-fold structure originates in the body of
the SMC and extends approximately 4 kpc towards the LMC (Muller et
al. 2003). It has clearly been significant in the evolution of the Bridge
and is the subject of a future work. The grey scale is a linear transfer
function as shown in the wedge and units are K$^.$arcsec . As this map is
integrated along the x image axis, the labeled Declination axis is correct
only for RA=2\hr08\mn8\secs.}
\epsscale{1}
\end{figure*}

Fig.~\ref{f:himap} shows an RA-Dec. peak intensity image of the \hi\
spectral line data cube used in this study. The diverse range of
morphologies in the \hi\ of the MB is striking in this image: some
regions are bright and filamentary, particularly in the southern
parts, while others are more tenuous and smooth. Fig.~\ref{f:vel-dec}
shows a peak pixel image of the same dataset viewed in the
Dec.-Velocity projection. The distribution of \hi\ in this figure
appears well-organised into at least two components. Notably, these
two components appear to be distinct also in declination. The velocity
separation of these components is $\sim$ 40 \kms.

The rapid spatial and velocity variability of the \hi\ in the MB shown
in Figs.~\ref{f:himap} \& \ref{f:vel-dec}, indicates a turbulent and
highly inhomogeneous morphology over the entire range of observed
spatial and velocity scales. Consequently, we have applied a separate SPS
analysis on four subregions within
the data set, as shown in Fig.~\ref{f:himap} with the overlaid grid.
Each region spans $\sim2.2\times2.2$ kpc (256$\times$256 pixels). We
will refer to these regions as (clockwise starting from the bottom
left on Fig.~\ref{f:himap}): the South East (SE), the North East (NE),
the North West (NW) and the South West (SW).

The SW region is dominated by bright, filamentary features across its
entire extent. The \hi\ in the SE region is similarly filamentary,
although the spatial coverage of the brightest feature is less
complete. The NE region defines an area of less diverse structure
where the \hi\ appears relatively tenuous and smooth.  The NW region
appears to be dominated by part of a large loop-like
feature. Fig.~\ref{f:vel-dec} shows that the northern two sub-regions
(NE and NW) encompass \hi\ which occurs at lower Heliocentric
velocities.  The SPS analyses was applied on each region separately.

\begin{figure*}
\centerline{\plotone{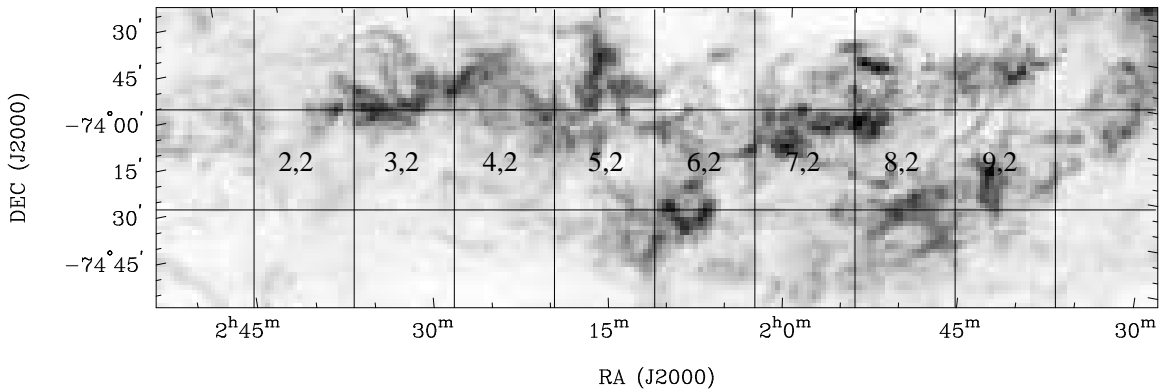}}
\caption{\label{f:scf_region}Peak pixel map of a subregion
  of the MB \hi\ dataset. The overlaid grid shows the eight regions for
  which the $S$({\bf $\Delta$r}) maps are calculated.  Each square labeled
  with a grid reference forms the centre of a larger square region over
  which the SCF is calculated.  The surrounding regions constitute the
  buffer. This entire area approximately corresponds to regions SE and SW of
  Fig.~\ref{f:himap}}
\end{figure*}

Due to the unpredictable response of the SCF to noisy data, we have
restricted the SCF analysis to a the brightest region of the Magellanic Bridge, shown
in Fig.~\ref{f:scf_region}. Much of this area has a spectral quality of $Q >
6$ (see Eq.~\ref{eq:quality} in Section~\ref{sec:scf_derivation}) . Restricting the tested data to  those with a
quality of this level limits the effects of noise on the SCF to $< 20\%$.
In contrast, the northern regions of the data set is more dominated by
noise, having areas where $Q < 6$.  Although binning can increase the
spectral quality, the degree of averaging necessary to effect a substantial
increase in $Q$ ultimately obscures the velocity structure, thereby
compromising the utility of the SCF. The
10$\times$3 grid overlaid on Fig.~\ref{f:scf_region} shows the eight
areas over which the SCF was averaged (see \S\ref{sec:calculating
scf}.  Each small square defines a 500 $\times$500 pc area ($63\times
63$ pixels).  As the telescope beam subtends approximately three image
pixels, a new data cube was generated that sampled every third pixel
in the $x$ and $y$ directions to ensure that the SCF samples independent
spectra.

\section{The Spatial Power Spectrum Analyses}
\label{sec:sps}

\subsection{Derivation of the SPS}

The two-dimensional SPS used here is derived following methodologies
used in previous studies (Crovisier \& Dickey 1983; Green 1993;
Stanimirovi\'c et al. 1999a; Stanimirovi\'c \& Lazarian 2001;
Elmegreen, Kim, \& Staveley-Smith 2001; Dickey et al.  2001).  We
start with two-dimensional images of the \hi\ intensity distribution.
These images are tapered with a Gaussian function to suppress the
$\sin(k)/k$ ringing that appears from the sharp edges of the original
image under the Fourier transform.  The power (square of the modulus
of the Fourier transform) is then measured as a function of wave
number (or spatial scale).  The power is then azimuthally averaged in
rings of logarithmically increasing intervals of wave numbers.

The interferometric (ATCA) part of this dataset has negligible
non-circularity of the $u-v$ sampling function (see also Muller et
al. 2003, Fig.~1) and no compensation was made for any ellipticity of
the $u-v$ sampling function. The median \hi\ column density for the
final \hi\ data cube is quite low ($\sim$4$\times$10$^{20}$
cm$^{-2}$), so any corrections for self-absorption are unnecessary
(see discussion by Minter 2002).

We emphasise that this dataset is a combination of observations
obtained with the Parkes single-dish telescope and with the ATCA
interferometer.  Therefore, a \emph{complete} range of spatial
frequencies was sampled from zero spacing, up to $\sim600 \lambda$,
corresponding to a minimum spatial scale of $\sim 100$ pc (note that
this is the \emph{contiguously sampled} range of spatial frequencies;
although spatial frequencies up to $\sim1600 \lambda$ have been
sampled, these are not part of a complete contiguous spatial frequency
distribution).

\begin{figure*}
\plotone{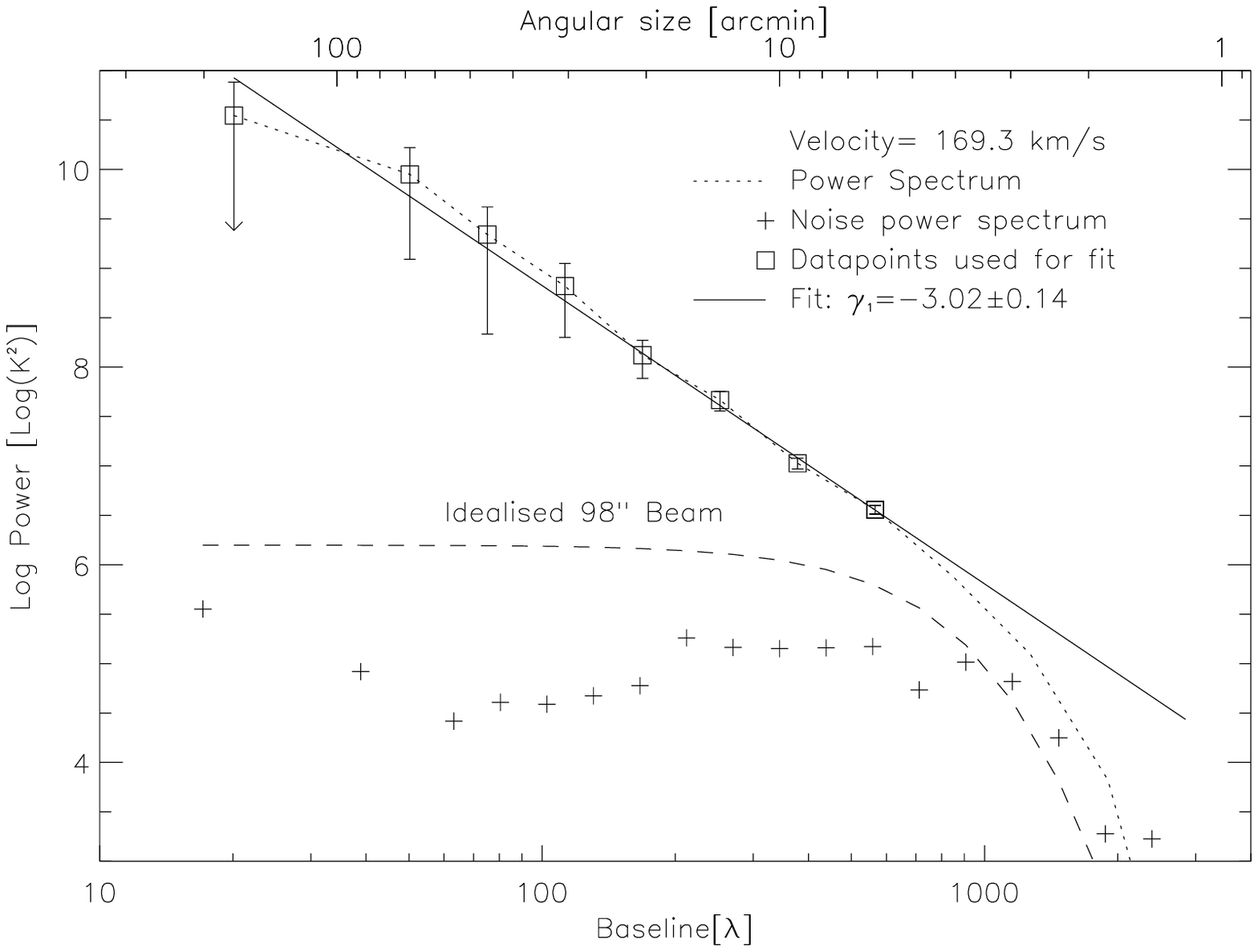}
\caption{\label{f:powerspect} Sample SPS from the South East
  sub-region (dotted line), at velocity $\sim$169 \kms\ (heliocentric). The
  spectrum is typical for channels which contain bright \hi. The fitted line
  shown (solid line) is an error-weighted fit, derived using the points
  marked with squares. Importantly, we see that the power spectra of the
  Bridge are able to be well fitted with a single component power-law.  Also
  shown on this plot is the power spectrum for a spectral channel containing
  only noise (crosses) and for the 98$\arcsec$\ beam (dashed line). The error
  bars mark one standard deviation, those showing a downward pointing arrow
  indicate a large lower error limit.}
\end{figure*}

A sample SPS from the MB dataset, typical for velocity channels with
significant \hi\ emission, is shown in Fig.~\ref{f:powerspect}.  Data points
calculated from a continuous range of spatial frequencies $<600 \lambda$
were used for fitting and are denoted by squares. The error bars in the
figure correspond to 1-$\sigma$ statistical uncertainties, which were taken
into account during the fitting procedure. 

At spatial frequencies \gesim700$
\lambda$ the effect of the ATCA beam becomes prominent, as shown with a
dashed line in Fig.~\ref{f:powerspect}. The SPS derived from an
emission-free velocity channel is shown in the same figure with crosses. We
see that the contribution of power to the SPS from the noise component of
the \hi\ dataset is negligible at low
spatial frequencies (by a factor of 10$^{5}$/10$^{6.5}\approx$0.03 at
$\approx$700$\lambda$) and is important only at the largest spatial
frequencies ($>700 \lambda$). As these frequencies are usually compromised by the ATCA beam, they are
already excluded from the fit.  
Although some efforts have been made to compensate for the effects of
noise and beam shape in power spectra (Green 1993), we have minimized these
effects by restricting the spatial frequency range to $<700 \lambda$
from the fit.

\subsection{The SPS as a function of velocity}

\begin{figure*}
\plotone{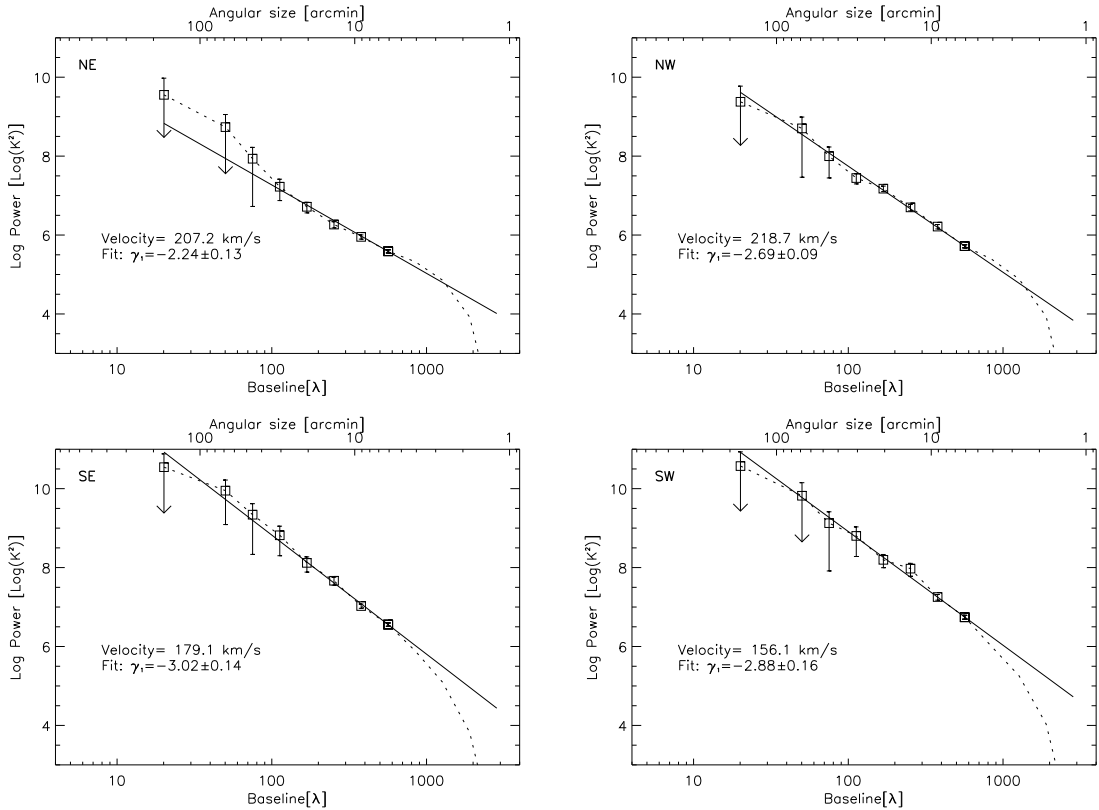}
\caption{\label{f:spectvel} Sample SPS from each of the
  four regions (marked in upper left) shown in Fig.~\ref{f:himap}.  These
  spectra are typically well fitted by a single component power-law. Error
  bars indicate one standard deviation and those showing downward pointing
  arrows indicate a large lower error limit.}
\end{figure*}

Fig.~\ref{f:spectvel} shows a typical SPS for all four sub-regions,
derived from single velocity channels with significant \hi\ emission.
A single power-law function, $P(k) \propto k^{\gamma}$, can be fitted
to the SPS for most of the spatial frequency range.  Similarly
featureless SPS were found for the SMC (e.g.  Stanimirovi\'c et
al. 1999a,b) and for the Galaxy (Green 1993; Dickey et al. 2001).

\begin{figure*}
\plotone{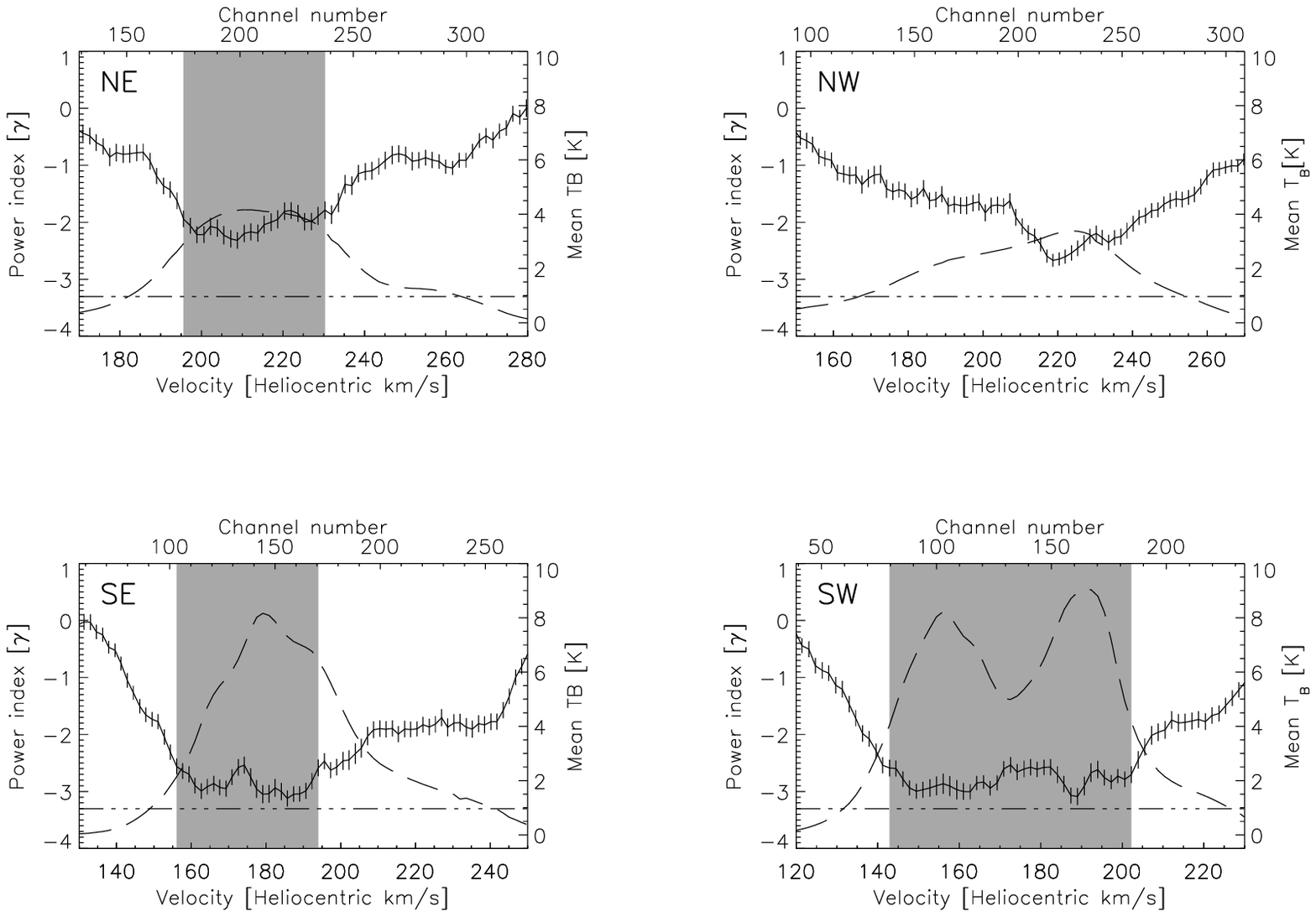}
\caption{\label{f:gamvel}Variation of the index $\gamma$
  (P(k)$\propto$k$^\gamma$) as a function of velocity for all four regions
  (marked in upper left) is plotted here as the solid line with error bars
  (\emph{left axis}). The total integrated \hi\ brightness is over plotted
  as the dotted line (\emph{right axis}). The shaded regions are those over
  which the power index is relatively constant and are the subjects for the
  study of the variation of spatial power-law index with integrated velocity
  thickness in Section~\ref{sec:gammadv}. The velocity intervals shown here
  are those for which $\gamma$\caplesim0 i.e. those channels which are not
  dominated by noise. The horizontal dot-dashed line marks the RMS T$_B$ for
  each subregion, as measured in line-free parts of the cube.}
\end{figure*}

The variation of the power-law slope with heliocentric velocity
$\gamma(v,\Delta v=1.6$ \kms$)$ for each of the four regions defined in
Fig.~\ref{f:himap} is shown in Fig.~\ref{f:gamvel}.  The left axis corresponds
to the power-law index $\gamma(v)$, while the right axis shows the
mean integrated \hi\ brightness temperature for a given velocity
channel.

The power-law index generally varies significantly with velocity, even
over the velocity channels with the brightest \hi\ emission.  The
nature of this variation is different in the different quadrants. This large
variation in $\gamma(v)$ contrasts to that which was found for the \hi\ in
the relatively homogeneous SMC: Stanimirovi\'c et al. (1999a)
measured no dramatic or large changes in the power-law index across the
entire ($\sim$200 \kms) velocity range. This has
important ramifications in the next part of the SPS analysis, where
the fundamental assumption is that the structure of the set under
study is largely homogeneous throughout the velocity range of the data.

To some extent, the plots of Fig.~\ref{f:gamvel} suggest a relationship between the mean brightness temperature and the index of the
power-law, in that a hight mean T$_B$ is associated with a steeper power
spectrum. This is can be seen for velocity channels
where the mean $T_{B}<$2K or so. This effect is primarily a
manifestation of the increasing relative importance of the noise
component: for channels where T$_B$ is low, the spectrum becomes
'whiter' (i.e. the spectrum flattens). 

In the case of the SW region, the variation of T$_B$ is very
large and we see that the variations of T$_B$ are not reflected by similar
variations of $\gamma$, which appears to have a limiting slope of
approximately $-$3. This is close to that expected from a Kolmogorov
flow.

\subsection{The SPS as a function of velocity slice thickness}
\label{sec:gammadv} 
We now investigate the extent to which velocity fluctuations affect
the apparent SPS of the \hi\ intensity distribution by applying the
VCA technique (Lazarian \& Pogosyan 2000) to \hi\ data cube subsets
averaged over a range of velocity intervals, which is just the
measurement of $\gamma(\Delta v)$.  As the VCA technique assumes that
the morphology of the original dataset is quite homogeneous over the
velocity range of interest, we will apply the VCA analysis only over
selected contiguous velocity windows for which the SPS index is
approximately constant to facilitate comparison with theoretical
models. In doing so, we avoid mixing kinematically different regions.
Unfortunately, choosing velocity windows limits the available velocity
thickness range considerably.

Velocity windows used for this analysis are shown as shaded bands in
Fig.~\ref{f:gamvel}. We selected regions with roughly constant
$\gamma(v,\Delta v=1.6\ \kms)$ and mean \hi\ brightness temperature
($\gtrsim$ 4 K) as a sample of relatively homogeneous regions. There is no usefully large velocity range over which the power-law
index is particularly constant in the NW region, so we do not attempt
to measure $\gamma(\Delta v)$ in this subregion.

Quantitatively, the ranges selected for each region follow:\\
\noindent {North West:} None\\
\noindent {North East:} 196 -- 230 \kms\\
\noindent {South West:} 143 -- 203 \kms\\
\noindent {South East:} 156 -- 194 \kms\\

\begin{figure*}
\plotone{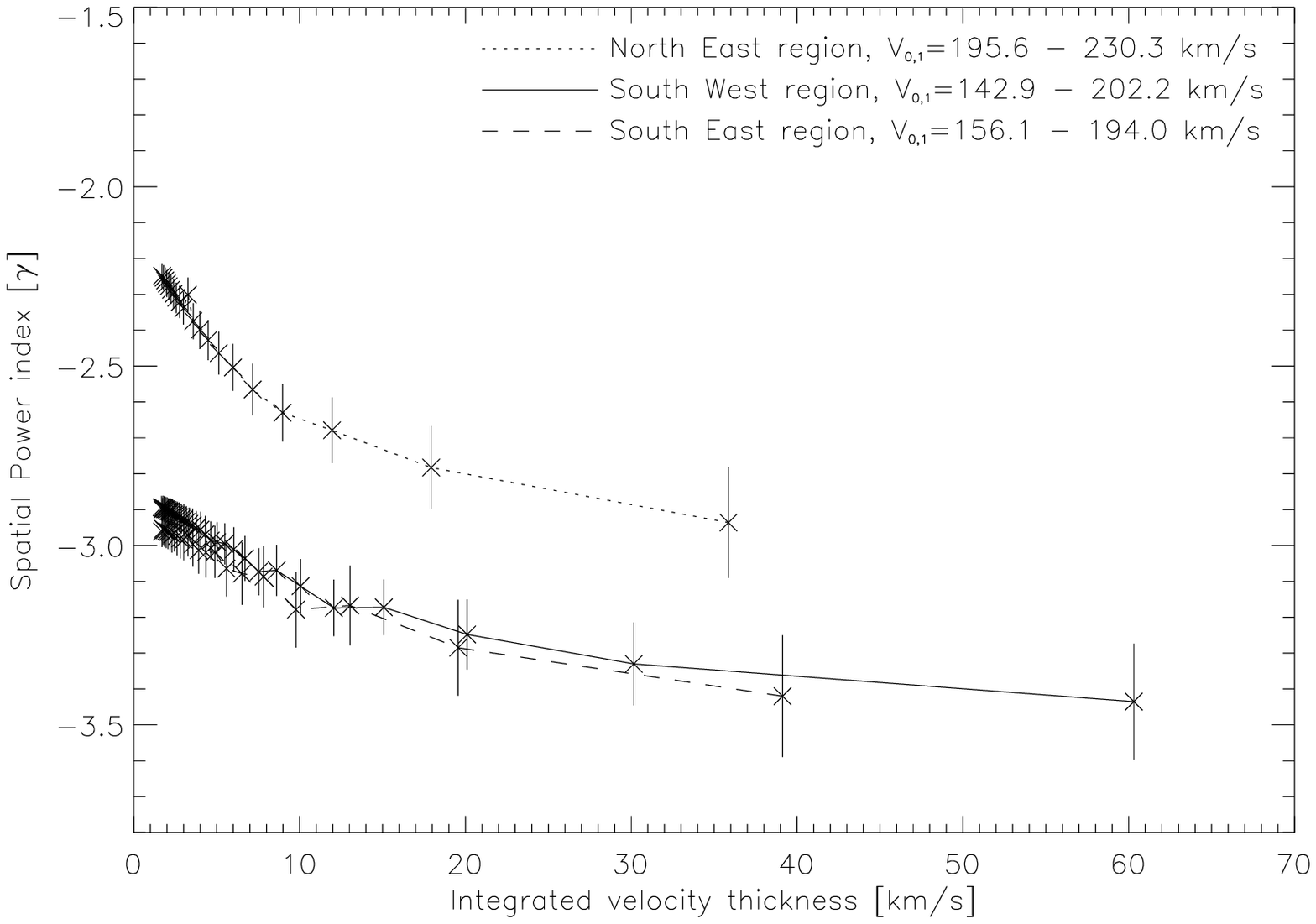}
\caption{\label{f:gamdv}Modification to the
  SPS index as a result of successively thicker velocity averaging for three
  of the four subregions shown in Fig.~\ref{f:himap}. The SPS index of the
  North East region is quite distinct from the two southern regions and
  changes more rapidly at narrower velocity intervals than does the southern
  regions.}
\end{figure*}

The variation of the SPS power-law index as a function of velocity
slice thickness for the three regions is shown in
Fig.~\ref{f:gamdv}. Generally, $\gamma(\Delta v)$ becomes steeper with
increasing velocity thickness for all three regions.  
This trend agrees with predictions by Lazarian \& Pogosyan (2000) and
suggests that both density and velocity fluctuations contribute to the measured
\hi\ intensity fluctuations for thin slices.  The selected range of
velocity channels for the SE region is narrower than that of the SW
region; however over the common range, $\gamma(\Delta v)$ is very
similar. This suggests that turbulence has comparable characteristics
in the two southern regions although, the power-law index derived for the NE
region appears very different. Over the whole range of $\Delta v$ the slope
for the NE region is significantly shallower than the slope of the southern
regions.

It is interesting to note that for all three 
regions $\gamma(\Delta v)$ starts to
converge a constant value around $\Delta v \sim 10$ \kms.
Lazarian \& Pogosyan (2000) predicted
that the presence of gas at different temperatures will show a 
characteristic `turn-over' in $\gamma(\Delta v)$. 
This slight `turn-over' of $\gamma$, seen in Fig. 6, around
$\Delta v \sim 10$ \kms could be an indication of a significant
contribution from warm gas with temperature $<10^{3}$ K.

In contrast, we do not see a similar converging trend for small
values of $\Delta v$ end of this plot. Lazarian \& Pogosyan (2000)
suggests that typically for cold \hi\ $\Delta v \sim 2.6$\kms\ would
be enough to approach the asymptotic value for `thin' slices (our
thinest channel is 1.65 \kms\ wide).  
Together with the observed turn-over at $\Delta v \sim 10$, the lack of convergence for
small $\Delta v$ suggests that \hi\ in the MB is
primarily in the warm neutral phase. However, absorption observations of a number of
continuum sources throughout the MB made by Kobulnicky \& Dickey (1999), have
revealed the presence of cold ($<20 K$) clouds. This cold phase is probably only a
small fraction of the MB ISM and as such, the mechanisms responsible for preserving
the cold component in equilibrium with the warmer ambient ISM remain to be
explained. 

\subsection{Spatial power as a function of azimuthal angle}

By examining the power as a function of azimuthal angle, we can probe for any
structure anisotropy of the MB \hi\ dataset.  Any such anisotropies will
manifest in the Fourier plane as significant structure as a
function of phase, with excess power aligned aproximately in the direction
perpendicular to that of the structure bias in the un-transformed image.
Such azimuthal anisotropy could result from
directionally-specific peturbations, for example, the effects of a
magnetic field (Esquivel et al. 2003), which may act to
organise the ISM by shaping the ionised or dusty component.

\begin{figure*}
\plotone{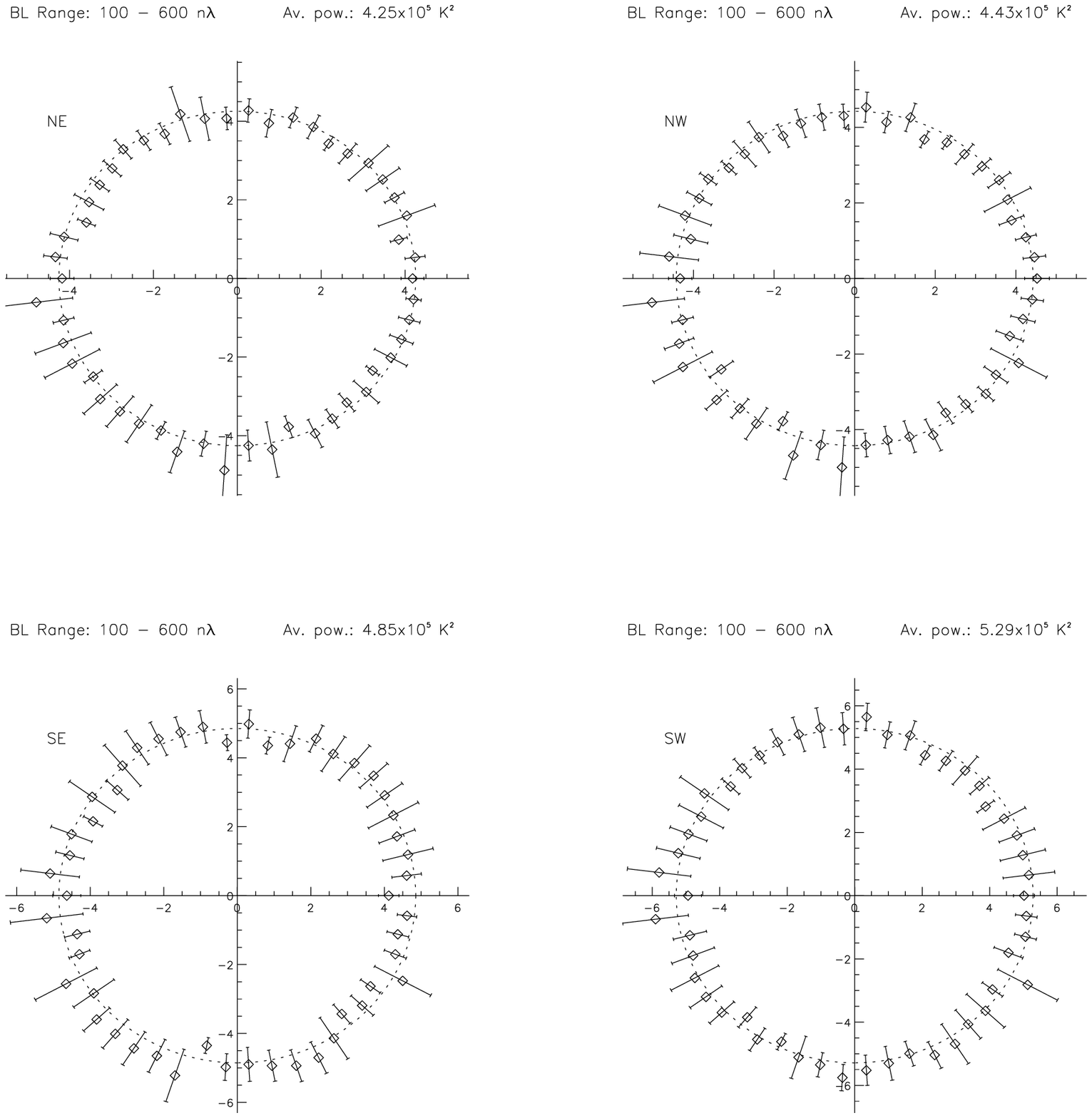}
\caption{The variation of power
  in the Fourier transform of an integrated intensity map for each of the
  four regions is plotted here as diamonds. The log of Power is shown as a
  function of azimuthal angle.  The dotted circle shows the mean power for
  this baseline interval. Error bars of one standard deviation show that
  there is no convincing consistent deviation from the dotted circle, and no
  indication of coherent structure in the Fourier transformed data for any
  of the four regions.} \label{f:power_azimuth}
\end{figure*}

Fig.~\ref{f:power_azimuth} shows the mean power as a function of
azimuthal angle across spatial scales 100-600$\lambda$ of the Fourier
transform for the \hi\ integrated intensity distribution of each of
the four subregions.  We find no evidence for any azimuthal anisotropy
in the Fourier plane, despite the filamentary and apparently east-west
orientation appearance of the \hi\ distribution in the southern
regions shown in Fig.~\ref{f:himap}.

It is likely that the dynamic range of any structure in the \hi\
images is insufficient to manifest as features in the Fourier
transform: for such a feature to be distinct it is necessary to have a
power at least 2--3 orders of magnitude larger than the surrounding
structure.  This is equivalent to an integrated temperature difference
of 10-30 K$^.$\kms. The apparent East-West structure of Fig.~\ref{f:himap} shows an
integrated brightness range significantly smaller than this, typically
only $\sim$2--5K$^.$\kms\ less than the ambient and less structured
\hi. In addition, these structures are not particularly well defined,
which will dilute the overall effect.

\section{The Spectral Correlation Function Analyses}
\label{sec:calculating scf}


The motivation for the use of the SCF here is intended as a benchmark, by
which we can compare the outcomes of the SCF and the SPS on the same
dataset, and to test the response of the SCF on a purely tidally-generated
system and in comparision to previous SCF studies of the LMC (Padoan et al, 2001).

\subsection{Derivation of the SCF}\label{sec:scf_derivation}

The algorithm used here to calculate the SCF is that employed by
Padoan et al. (2001) for the study of the \hi\ intensity distribution
in the LMC. The SCF is defined in the following way (see also
Rosolowsky et al. 1999; Padoan, Rosolowsky, \& Goodman, 2001):
\begin{equation}
  S_0({\bf r},{\bf \Delta r})=\biggl(\frac{\mathcal{S}_0({\bf r},{\bf \Delta
      r})}{\mathcal{S}_{0,N}(\bf r)}\biggr) \label{scf_eq}
\end{equation}
with the numerator being defined by:
\[ \mathcal{S}_0({\bf r,\Delta r})=\biggl(1-
\sqrt{\frac{\sum_{v}[T({\bf r},v)-T ({\bf r+\Delta r},v)]^2}{\sum _{v}T({\bf
      r},v)^2+\sum_{v}T ({\bf r+\Delta r},v)^2}}\biggr)\] 
The noise term in the denominator is defined for the subject spectrum,
$T(\mathbf{r},v)$, by:
\[  \mathcal{S}_{0,N}({\bf r})=1-\frac{1}{Q({\bf r})}\]
where $Q({\bf r})$ is the spectrum `quality' and is given by:
\begin{equation}
\label{eq:quality}
Q(\mathbf{r}) = \frac{1}{R}\sqrt{\sum_{v}T^2(\mathbf{r}, v)}
\end{equation}

and $R$ is the RMS in signal free parts of the cube.

We aim to generate maps of the spatial variation of $S_0$({\bf
r,$\Delta$r}), for a number of regions in the \hi\ MB dataset. The
algorithm used is outlined below:

\begin{enumerate}
\item \emph{Dataset Selection.} A square region of a side with $N$ pixels is
  defined within the data cube. The SCF is calculated iteratively using
  every spectrum in this region in turn as the subject spectrum.

\item \emph{Calculate $S_0(\mathbf{r,\Delta r})$ map.} A new larger
region of an area of $(2N-1)^2$ is defined with the subject spectrum
at the centre.  The $S_0$({\bf r,$\Delta$r}) for each spectrum in this
region is calculated relative to the centre subject pixel. This
process forms a two dimensional map, where the value at each point is
proportional to the similarity of the spectrum to that of the centre
spectrum. The value for the centre pixel in this SCF map is unity and
all other pixel values are $\leq$1. This map is stored for subsequent
averaging.

\item \emph{Iteration and averaging.} Step 2 from above is repeated for each
  pixel in the initial $N\times N$ region of interest. Ultimately, a total
  of $N^2$ maps are obtained, each having an area of $(2N-1)^2$. The final
  result is determined by finding the mean for all of the calculated SCF
  maps: $S$({\bf $\Delta$r})= $\bigl<S_0$({\bf
    r,$\Delta$r})$\bigr>_{\mathbf{r}}$
\end{enumerate}

This approach requires a `buffer' of spectra, at least $N$ pixels wide
around the region of interest.  Without this buffer, there is not a
uniform sampling population for each pixel of the SCF map and edge
effects become important.  The numbering in Fig.~\ref{f:scf_region}
shows the grid reference of the centre square region defining the
subject spectra (step 1 above). $S_0$({\bf r,$\Delta$r}) is calculated
for all spectra within each box relative to all spectra within 20
pixels (step 2 above) and then averaged over all $21^2$ pixels in that
map (step 3). Overall, we obtain 8 different maps of
$S_0$({\bf$\Delta$r}) from this dataset.

\subsection{Results of the SCF analyses}

\begin{figure*}
\centerline{\plottwo{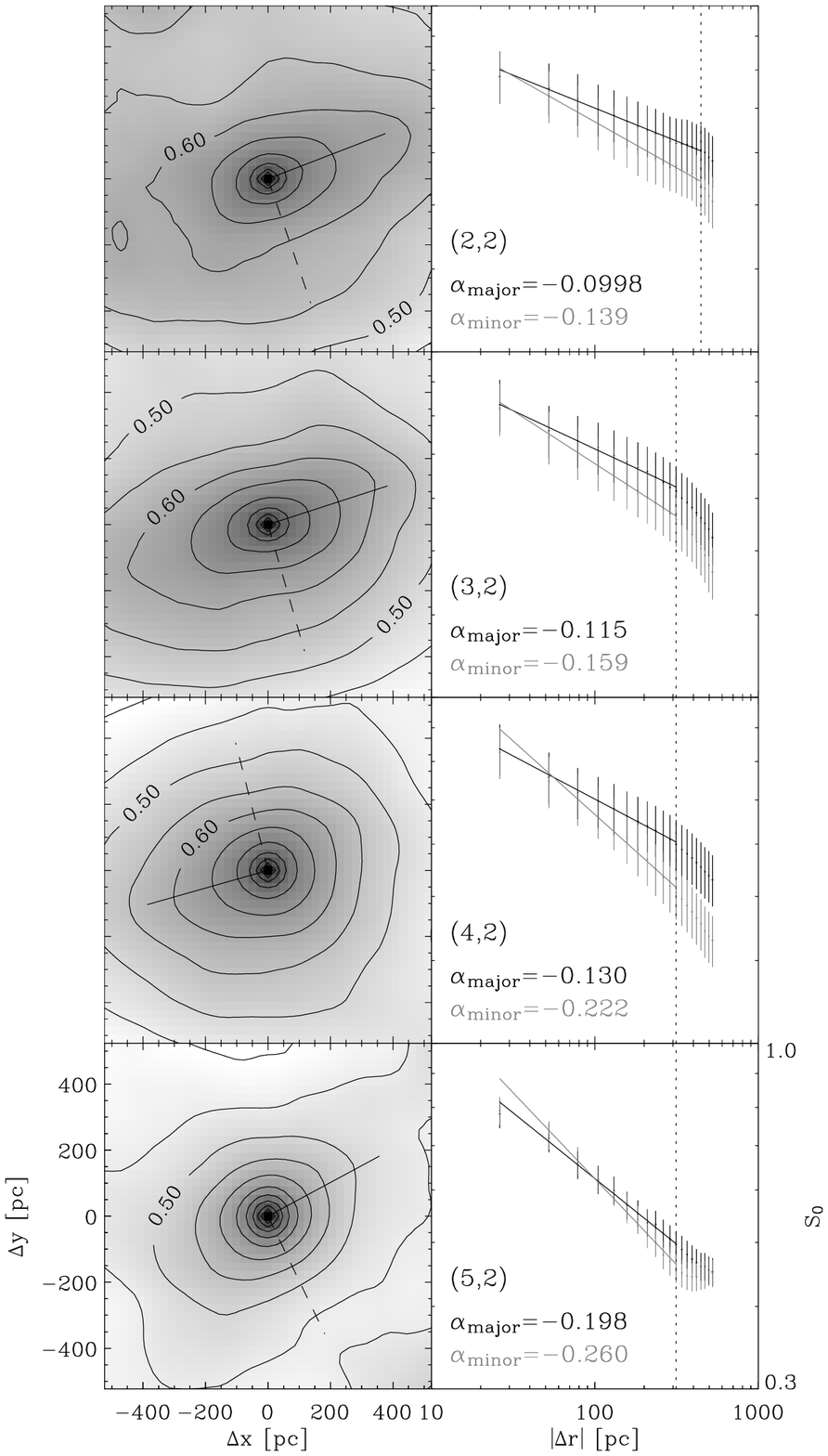}{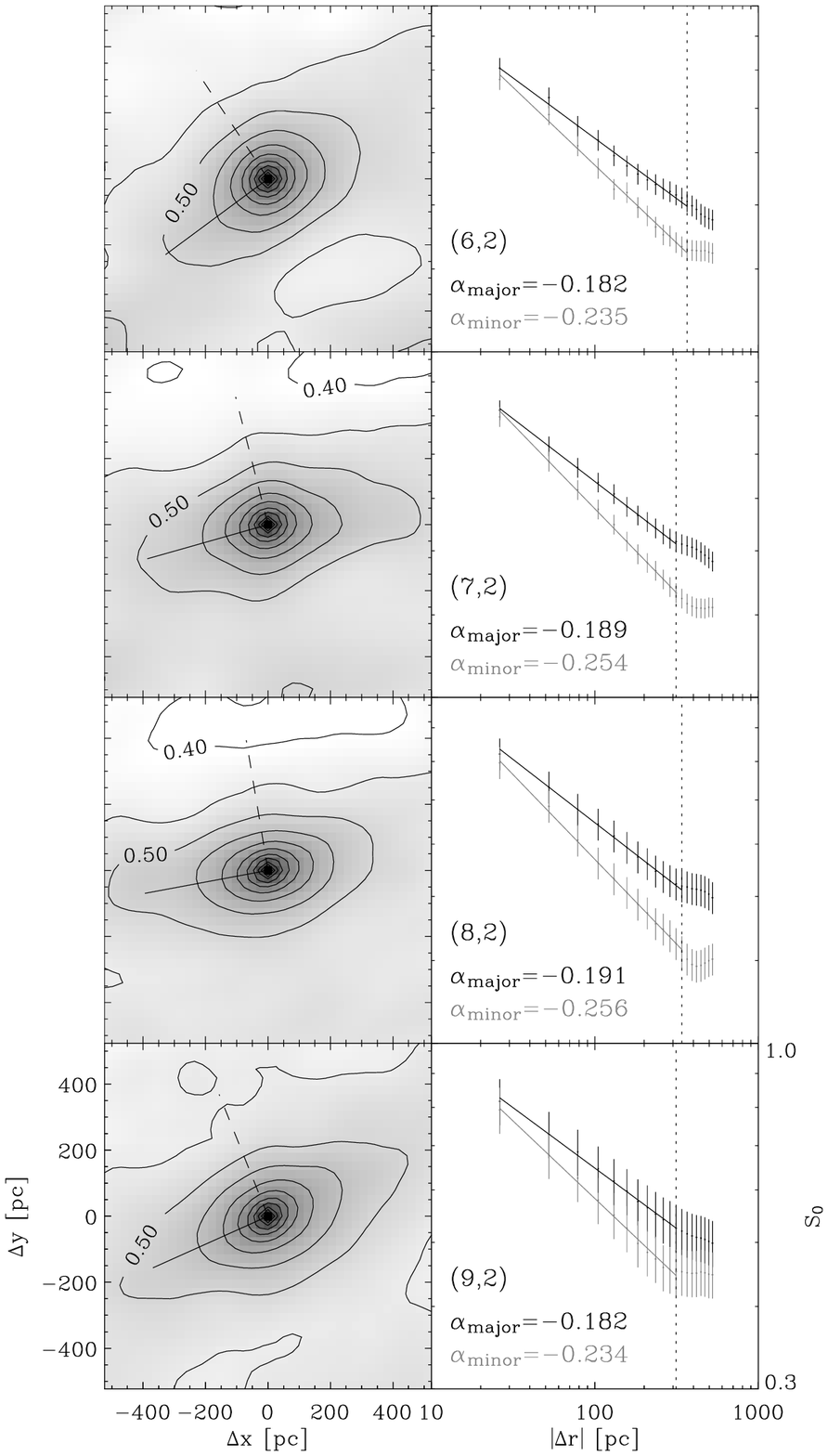}}
\caption{\label{pixel_map}
  Results of SCF analysis on the \hi\ in the MB for grids (2,2)--(5,2) (left panel; see
  Fig.~\ref{f:scf_region}) and for
  grids (6,2)--(9,2) (right panel). The left part of each panel shows the map of
  $S_0(\mathbf{\Delta r})$ according to the SCF algorithm in the text. The
  right part of each panel shows 
  radial plots of $S_0(|\mathbf{\Delta r}|)$.  The two cuts indicated in the left panel (solid and dashed line)
  appear in the right panel for the major (black) and minor (gray) axes of
  the elliptical $S_0$ map respectively. We see that without exception, the
  ellipses are aligned with the major axis (i.e. the more slowly changing
  axis) approximately East-West. Each panel is annotated to
  indicate the corresponding \hi\ subset defined on
  Fig.~\ref{f:scf_region}.}
\end{figure*}

The results for the SCF analysis are shown Fig.~\ref{pixel_map}.  The
left-hand panels show the eight SCF maps, corresponding to each of the
labeled regions in Fig.~\ref{f:scf_region}. Note that the maps
contain significant structure, so that simply calculating $S_0$ as a
function of $|\mathbf{\Delta r}|=\sqrt{x^2+y^2}$ (i.e. the
azimuthally-averaged approach
taken by Padoan et al.  2001) does not characterise the
$S_0(\mathbf{\Delta r})$ appropriately.  As such, the right-hand
panels of Fig.~\ref{pixel_map} show the behaviour of
$S_0(|\mathbf{\Delta r}|)$ along the major and minor axes of the
elliptical contours. The map is clearly anisotropic and indicates some
structure bias present in the organisation of the gas in the MB. In
particular, the major axes of the ellipses appear to be generally
aligned with direction of filamentary structure in the MB.

In order to interpret the SCF maps, we have fitted an elliptical power-law
function to the distribution.  We modeled the typical $S_0$ surface using a
function with parameters $c_i$:
\begin{eqnarray}
  f(x,y) &=& c_0 (x^2+y^2)^{\beta/2}\mbox{, where}\\ \beta &=& c_1
  \cos^2\left(\phi-c_3\right)+c_2 \sin^2\left(\phi-c_3\right),\ c_1 \geq c_2
  \mbox{, and} \nonumber\\ \phi & = & \arctan(y/x) \nonumber
\label{fit_fcn}
\end{eqnarray}

This function characterises the $S_0$ surface with two separate
power-laws falling off in orthogonal directions from $\mathbf{\Delta
r}=0$ with indices $c_1$ (major axis) and $c_2$ (minor axis)
respectively. The angle $c_3$ represents the orientation of these axes
with respect to the coordinate axes.  The data $S_0(x,y)$ are weighted
inversely proportional to their distance from $S_0(0,0)$. The
singularity at $f(0,0)$ for negative values of $c_1$ and $c_2$ is
ignored.  We fit the function $f$ to $S_0$ for every position that is
completely sampled.

Padoan, Goodman, \& Juvela (2003) point out a relationship between the
power-law scaling of $S_0(|\mathbf{\Delta r}|)$ and the line width of
the spectra considered.  We find no such correlation present in the MB
\hi\ data, however, the $^{13}$CO molecular cloud data used by Padoan,
Goodman, \& Juvela (2003) are systems with well defined size--line
width relationships and where such line profiles can typically be
described with only a single component. In contrast, the \hi\ data
used here are complex, and show multiple velocity components for any
line-of-sight (e.g.  McGee \& Newton, 1986; Muller et al. 2003)

\begin{figure*}
\epsscale{0.6}
\plotone{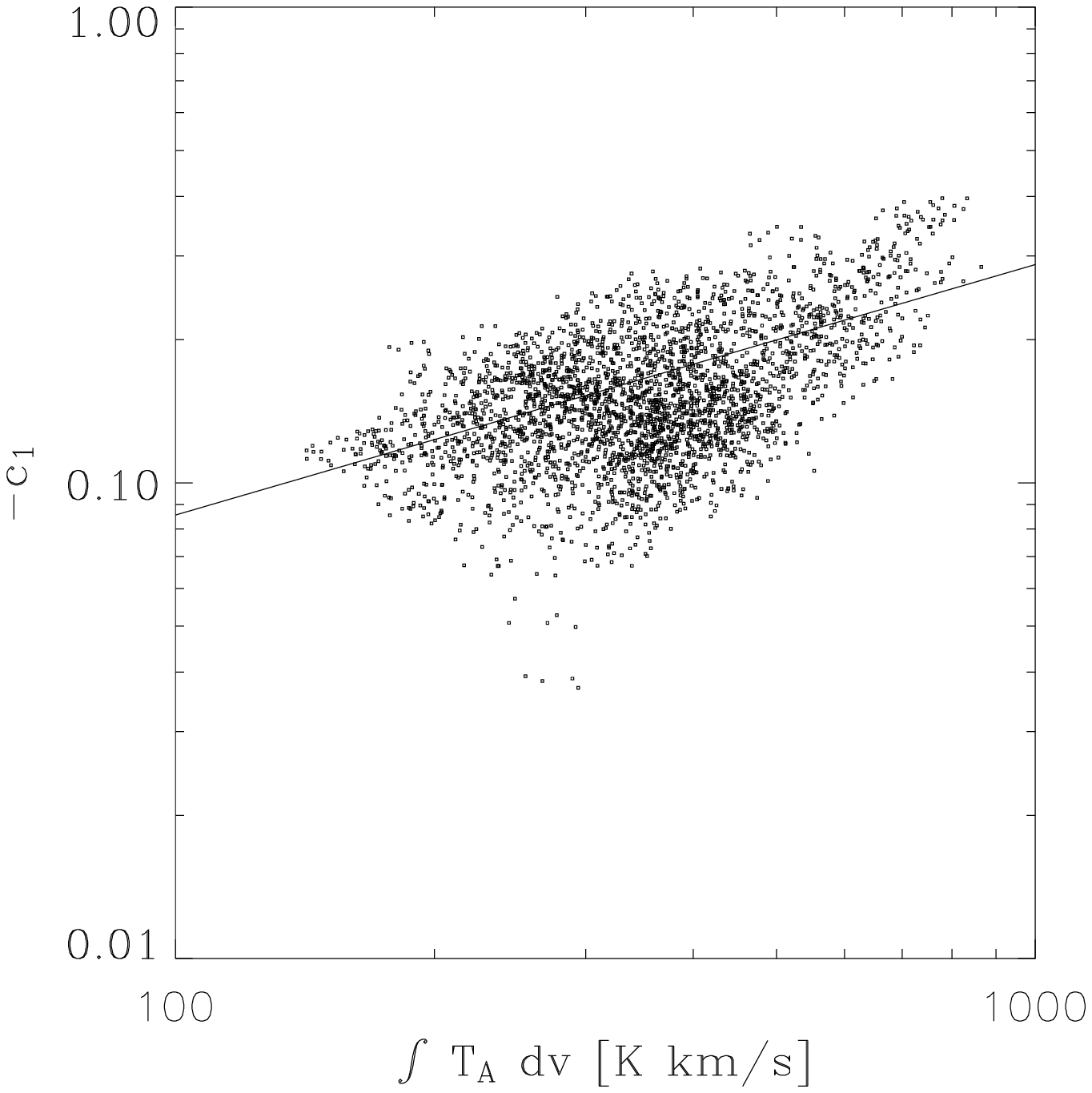}
  \caption{\label{int_scf} Correlation between
    integrated intensity and the major axis power-law index $|c_1|$ for the
    SCF.  The correlation includes only those pixels sufficiently far from
    the edge of the map to avoid edge effects.  The solid line plots the
    best fitting correlation $|c_1|\propto (\int T_A dv)^{0.5\pm 0.1}$.}
\epsscale{1}
\end{figure*}

There is evidence for a correlation between the power-law index of the SCF
and the integrated intensity. This is most notable when step 3 of the SCF
processing is omitted and the major and minor axes are derived by fitting
without averaging.  Fig.~\ref{int_scf} shows this correlation with respect
to the major axis index, $c_1$, for all points outside the 21 pixel buffer
at the edge of the map. The correlation is not outstanding, but appears to
obey a power-law such that $|c_1|\propto (\int T_A dv)^{0.5\pm 0.1}$.
Padoan, Rosolowsky, \& Goodman (2001) note that higher density gas appears to
be traced by a steeper power-law slope. However, this effect may be due to
different spatial distributions for different molecular tracers.

\section{Discussion of Statistical results}
\label{sec:sps_disc}

\subsection{Spatial Power Spectrum}

We find that the SPS of the \hi\ intensity distribution in the MB can
be well described by a single power-law function over a wide range of
spatial scales (29 pc to at least 2 kpc).  As in the case of the SMC
(Stanimirovi\'c et. al, 1999a,b), no significant changes to the
power-law shape of the SPS were found when varying heliocentric
velocity or the thickness of velocity channels, though the power-law
index varies significantly.  This indicates that the MB
is likely to have a large line-of-sight depth. In contrast, the SPS derived from the
\hi\ intensity distribution in the LMC showed a break at the spatial
scale equivalent to $\sim$100 pc (Elmegreen, Kim \& Staveley-Smith,
2001) which was interpreted as evidence for the finite thickness of
the LMC disk. However, the SMC and the MB have large-scale morphologies which are
distinctly different from the LMC, namely that they are not
approximately face-on systems and have a substantial line-of-sight
depth.

There is a clear variation of $\gamma(\Delta v)$ over the range of
velocity widths considered, and the NE section of the analysis has
qualitatively different variation than the SW and SE sections.
Lazarian \& Pogosyan (2000) relate the measured variations of
$\gamma(\Delta v)$ to physical properties of the turbulence, a
particular strength of SPS analysis.  In particular, $\gamma(\Delta
v)$ can be related to the indices of the three-dimensional power
spectra of the density and velocity fluctuations.  If the density
fluctuation power spectrum is a power law with index of $n$ and the
velocity fluctuation power spectrum has an index of $\mu$ the
relationships between $\gamma$, $n$ and $\mu$ are given in Table
\ref{tab:lazarian} for $n > -3$ (Shallow 3-D density) and $n < -3$
(Steep 3-D density).

 \begin{table*}
 \begin{center}
 \begin{tabular}{c|cc}\hline\hline
$\gamma$ &Shallow 3-D density& Steep 3-D density\\
\hline 
Thin slice ($\Delta v$ small) & ${n-\mu/2-3/2}$&${-9/2-\mu/2}$\\ 
Thick slice ($\Delta v$ large) & $n$&${-3/2+\mu/2}$\\ 
Very Thick Slice ($\Delta v$ very large)& $n$&$n$\\ 
\hline
 \end{tabular}
 \end{center}
 \caption{The value of the observed power-law index $\gamma$ for
 different values of $\Delta v$ in different turbulent regimes.  $n$
 is the static spatial power law index of the turbulence which is
 modified by velocity fluctuations with a slope of $\mu$ (Lazarian \&
 Pogosyan, 2000).}
 \label{tab:lazarian} 
\end{table*}

Following the description of Lazarian \& Pogosyan (2000): if we assume that the smallest $\Delta v$ traces `thin' velocity
slices and that the largest $\Delta v$ corresponds to `thick' velocity
slices, we can estimate the factor by which the velocity fluctuations modify the SPS
density spectrum.  The derived values for the slopes of the density
and velocity spectra are listed in Table~\ref{tab:3dvelocity}.  These
values are given as upper and lower limits since asymptotic solutions
were not fully met, especially at the lower $\Delta v$ end.  

Similarly, if we assume that the thickest velocity slices used in the study of
Section~\ref{sec:gammadv} are probing the pure density-dominated
regime of Lazarian \& Pogosyan (2000), then the 3-D density index for
the SW region has a lower limit of $n\approx -3.4$, indicating that
this region is in the \emph{velocity}-dominated regime. Using the SPS
index for the thinnest velocity slice of this region leads to a lower
limit estimate of the velocity fluctuations $\mu\approx -4$ (Table 2).
Similarly for the NE region, the
SPS index for thickest velocity slice has a lower limit of
$n\approx-2.9$, indicating that this region is in the
\emph{density}-dominated regime and that velocity fluctuations are
small in this region. The derived 3-D velocity index has an upper
limit of $\mu\sim-$4.5 and is extremely steep, indicating an
overwhelming deficiency of high-velocity fluctuations.

\begin{table*}
\begin{center}
\begin{tabular}{lccc}\\\hline\hline
  &{\large$\gamma$}$_{Thin}$& {\large$\gamma$}$_{Thick}$ ($=n$)&$\mu$ 
3-D Velocity\\ 

  &\emph{(lower limit)}&\emph{(upper limit)} &\emph{(upper limit)}\\
  &&3-D Density&\\\hline

  South East&$-$3.05$\pm$0.06&$-$3.4$\pm$0.2&$-$3.6$\pm$0.3\\ 
                                             
  South West&$-$2.94$\pm$0.04&$-$3.4$\pm$0.2&$-$3.9$\pm$0.2\\ 
                                             
  North East&$-$2.24$\pm$0.03&$-$2.9$\pm$0.2&$-$4.5$\pm$0.4\\ 
  \hline\hline

\end{tabular}
\end{center}
\caption{SPS indices for thickest and thinnest velocity slices for the
three sampled regions in the MB (SE, SW and NE regions of
Fig.~\ref{f:himap}). The degree by which two-dimensional velocity
fluctuations affect the SPS index are shown, along with the estimated
3-D velocity index (see Lazarian \& Pogosyan, 2000 and
Table~\ref{tab:lazarian}).}
\label{tab:3dvelocity}
\end{table*}

The steepening of the SPS slope with increasing integrated velocity
thickness, as predicted by Lazarian \& Pogosyan (2000), was also
measured from \hi\ in the SMC (Stanimirovi\'c \& Lazarian 2001) and for a
test region in the Galaxy (Dickey et al. 2001).
Table 2 shows that the
slopes for the southern two regions in the MB are generally similar to
those derived for the SMC which has a $n=-3.3$ and $\mu=-3.4$
(Stanimirovic \& Lazarian 2001).  Both the SMC and the MB appear to have
power-law indices that are more shallow that what is predicted for a
Kolmogorov turbulent spectrum.
For the case of \hi\ in the Galaxy
a shift of the SPS slope from about $-$3 to $-$4
after velocity integration was noticed by Dickey et al. (2001). 
Other studies of the LMC (Elmegreen, Kim \& Staveley-Smith 2001) 
and the Magellanic Stream (Stanimirovi\'c et al. 2002) also 
show apparent steepening of the
power spectra with increasing integrated velocity thickness.

A different interpretation of the power-law SPS in the case of the SMC
was provided by Goldman (2000) whereby a featureless SPS was
interpreted as a signature of turbulence being driven on spatial
scales larger than the size of the SMC.  This turbulence was induced
by local instabilities due to tidal interactions in the Magellanic
System. This approach also defines the relationship between the 2-d
intensity and 3-d density statistics, where the Power-law index of intensity
fluctuations, $\gamma$, is related to the power-law index of density
fluctuations, $n$, by $\gamma=n+1$.  In the MB case studied here, this
approach will result in $n\sim-4$ for the Southern two regions, and
$n\sim-3.2$ for the North-East region.

The second important result from these studies of the MB, is that the
slopes of the SPS derived for the northern and southern regions appear
to be strikingly different: the northern part shows a distinctly
shallower (although still monotonic) power spectrum, indicating a
relative lack of large-scale power (or equivalently, an excess of
small scale power). Furthermore, these results could be interpreted as
evidence for a warm ($\sim$10$^3$ K) component. This is slightly perplexing,
given the results of absorption detections
which indicate the presence for a very cool ($\sim$20 K) component.

Muller et al. (2003) have observed that at least two distinctly different
\hi\ velocity components exist throughout the MB, being centered on about 150
and 190 \kms\ (see also Fig.~\ref{f:vel-dec}).  Furthermore, the two
components are distinct in declination: the higher velocity component
is predominantly at northerly declinations and is sampled by the NE
and NW regions, while the lower velocity component is sampled by the
SE and SW regions.

Numerical simulations (Gardiner, Sawa \& Noguchi, 1994) suggest that
the two velocity components could be associated with arms extending
from the SMC. In this model, the more distant arm is projected as a
higher velocity and slightly northern feature, while the nearer arm
correlates with lower velocity component. The nearer arm is clearly
contiguous with the SMC and LMC may be more properly regarded as 'The
Magellanic Bridge'. The distances to these two arms have been
predicted by Gardiner, Sawa \& Noguchi (1994) to be centered at 50 and
70 kpc, with an expected line-of-sight depth for both entities of
$\sim$5--10 kpc. We propose that as an explanation for the
differences in the behavior of the power indices of the northern and
southern regions in Figs.~\ref{f:gamvel} \& \ref{f:gamdv}, these regions
encompass two separate and distinct arms of the SMC. We view the arms
projected approximately on top of one another, whereas these features are
not well spatially connected at all. 

\subsection{Spectral Correlation Function}

Like the SPS, the SCF shows a logarithmic dependence of scale which
can be parameterised with a single component.  We do not expect the
slopes of the SCF and SPS to be consistent, since the SPS probes
structure in excess of the mean, large scale power (i.e. that
represented by the zero lag), whereas the SCF algorithm, as used here
and by Padoan et al. (2000), does not factor for a mean brightness
level which will otherwise tend to flatten the SCF slope.  Interestingly, the SCF
slopes derived here for the MB dataset are largely consistent with
those derived from the \hi\ in the LMC ($S_0({\bf \Delta
r})\sim-$0.15 - $-$0.4; Padoan et al, 2001).

The lack of radial symmetry in the maps of $S_0$({\bf$\Delta$r}) is
itself an interesting result. It appears then that the
SCF is more successful in detecting structure trends over a smaller
dynamic range than the approach using the Fourier transform and the SPS in
Section~\ref{sec:sps}.

The obvious structure in the approximate East-West
direction (i.e. the line joining the SMC and LMC) indicates a more
persistent similarity of spectra along that direction. What we see
here is that the rate of change of the entire \emph{spectra} (rather
than a single velocity channel, or integrated brightness) across the
entire sampled velocity range is more gradual in the East-West
direction than in the North-South direction. 

We interpret the East-West $S_0$({\bf$\Delta$r}) structure as a result
of the tidal stretching imposed by the LMC-SMC interaction, although
in this case the morphology of the \hi\ is already visually suggestive
of this process.  We should also bear in mind that the brightest parts
of the dataset can dominate the SCF results, biasing steeper power-law
indices and the overall structure of the SCF maps. A floating
temperature scaling factor in the SCF algorithm would probably go some
way to removing this bias.

We note that there is {\it no} evidence for a break in the Bridge SCF
power law, as was found for the LMC at $\sim$100 pc,
although these two objects are widely separated in space and there is
no outstanding reason that they should share a common morphology at
that scale.  However, since a typical width of the structures in the
filaments is $\sim 100$ pc in the plane of the sky, and we do not see
a departure from a power law on this scale, this may be another
indication of significant line-of-sight depth in the Bridge.  We see
in Fig.~\ref{pixel_map} that there may be some evidence for a
consistent break in the power-law for both the major and the minor
axes, at $\sim$300---500 pc.  However, this departure occurs at scales
which are comparable to the maximum size of the tested areas
(i.e. $\sim$500 pc).  The degree to which the boundary and size of the
sample affects the SCF has not been well studied, and it is probable
that the departure is influenced by the edge of the sample region. As
such, any interpretation for the reasons of the observed departure at
this scales should be made with caution. It is not appropriate to
simply expand the spatial size of the sampled areas since this would
sample data which is excessively dominated by noise and the SCF
algorithm will be affected unpredictably.

In this study, we did not attempt to study the behaviour of the SCF in a
similar way to that used by the velocity-component analysis (VCA), namely,
measuring the modification to the SCF as a function of varying velocity
integrated width.  As the SCF operates in the image domain, we expect that
by averaging across the velocity fluctuating component the SCF will flatten
somewhat, and that there will be some correlation of the SCF slope with
velocity integrated width. However, to conduct this test
appropriately, a study with a well characterized dataset should
be used, rather than the complex \hi\ profiles of Magellanic Bridge dataset.

\section{Summary}
\label{sec:discussion}

We have attempted two statistical analyses of the filamentary \hi\ in the
Magellanic Bridge. Together with the fact that there also appears to be an
identifiable discontinuity in the velocity distribution of the \hi\ in the
MB, the most significant result from the calculation of the spatial power
spectrum (SPS) here has
provided support for a scenario suggested by numerical simulations, where the southern and northern parts of the Bridge represent the projection of
two distinct arms of gas emanating from the Small Magellanic Cloud (SMC). 
These findings suggest that we now need to
re-assess the current interpretation of the Bridge as a single filamentary
feature.

Similarly to the SMC, the power spectra throughout the 
MB are well described as a power law with a single power component.  
In particular, the southern region which contains brighter and more
turbulent \hi, have 3-d density and velocity indices similar to that which was
found for the SMC.
The more tenuous and higher velocity (i.e. northern) parts show a
significantly shallower SPS, with the 3-d density slope of $-2.9$. 
This is indicative of relatively quiescent gas. The
3-d velocity index ($-4.5$) 
of this region is significantly steeper that for the case of a
Kolmogorov turbulent flow.

We find that the a velocity component analysis (VCA) of the \hi\ in the MB behaves, on the whole, 
according to
predictions by Lazarian \& Pogosyan (2000), where modifications to the
power spectra by velocity fluctuations can be removed by integrating
over a velocity window. This has again highlighted the dissimilarities
between the northern and southern parts of the MB. The
contributions by velocity fluctuations to the southern part of the
MB are again similar to those in the SMC, whereas the SPS
from the northern part of the MB shows evidence for a very steep
three-dimensional velocity index and a lack of rapid velocity
fluctuations. 

A study of the azimuthal isotropy of the Fourier transform of the \hi\
dataset does not show any such East-West structure which is apparent
in the \hi\ peak and integrated intensity maps. It is likely that the
dynamic range of the Fourier transformed dataset is insufficient to
distinguish any such structure tends from the ambient \hi. Conversely,
the spectral correlation function (SCF), due to the nature of the algorithm, is much more successful
in indicating more subtle, local trends in structure.

The analysis using the SCF has shown more quantitatively the effects
of tidal stretching on this dataset. It confirms that the observed
\hi\ spectra in this region vary more slowly along the approximate
direction of the tidal stretching. Although limitations of the
application of the SCF has confined the region sampled to the
brightest parts of the MB, the SCF shows enormous potential as a tool
for more easily parameterising suspect tidal features, insofar as the
direction and extent of the tidal perturbation.

Although the behavior of the SCF cannot be directly related to the
physical properties of the underlying turbulence, the SCF provides
some provocative results.  We see that, like for the SPS, the SCF shows
a power-law dependence with indices compatible with previous studies
of the LMC \hi\ datset.  However, the results from the SCF analysis
also hint at a characteristic scale length, which is not observed in
the SPS of the same region and are not at scales consistent with the
studies of the LMC. It is important to bear in mind that the departure
observed here is at the limits of the sampled scale range and any
interpretation should be made carefully.  However, we can conclude
that the SCF and the SPS do not appear to measure structure variations
in the same way. Specifically, the SCF appears to be much more
sensitive to low-power and small scale variations in structure.  It
will be necessary to explicitly test and compare behaviors of the SCF
and SPS to different turbulence types before the full application of
the SCF can be recognized.

\section{Acknowledgments}
The Authors would like to thank Alex Lazarian for his time in reading the
draft and providing helpful suggestions for improvement. The Arecibo
Observatory is part of the National Astronomy and
Ionosphere Center, which is operated by Cornell University under a
cooperative agreement with the National Science Foundation. SS acknowledges
support by NSF grants AST-0097417 and AST 9981308.


\begin{thebibliography}
\expandafter\ifx\csname natexlab\endcsname\relax\def\natexlab#1{#1}\fi

\bibitem[Crovisier \&Dickey(1983)]{1983A&A...122..282C}Crovisier, J.,
  Dickey, J.~M.\ 1983, \aap, 122, 282

\bibitem[Dickey et al.(2001)]{2001ApJ...561..264D} Dickey, J.~M., 
McClure-Griffiths, N.~M., Stanimirovi{\' c}, S., Gaensler, B.~M., \& Green, 
A.~J.\ 2001, \apj, 561, 264 

\bibitem[Elmegreen, Kim, \& Staveley-Smith(2001)]{2001ApJ...548..749E} 
Elmegreen, B.~G., Kim, S., \& Staveley-Smith, L.\ 2001, \apj, 548, 749 

\bibitem[Elmegreen \& Scalo(2004)]{astro-ph/0404451} Elmegreen, B. G.,Scalo, J.,2004 astro-ph/0404451

\bibitem[Esquivel, Lazarian, Pogosyan, \& Cho(2003)]{2003MNRAS.342..325E} 
Esquivel, A., Lazarian, A., Pogosyan, D., \& Cho, J.\ 2003, \mnras, 342, 
325 

\bibitem[Gardiner \& Noguchi(1996)]{gn96} Gardiner,
 L.~T.~\& Noguchi, M.\ 1996, \mnras, 278, 191

\bibitem[Gardiner, Sawa, \& Fujimoto(1994)]{gsf94} Gardiner, L. T.,
Sawa, T., Fujimoto, M.\ 1994, \mnras, 266,567

\bibitem[Green(1993)]{1993MNRAS.262..327G} Green, D.~A.\ 1993, \mnras, 262, 
327 

\bibitem[Goldman(2000)]{2000ApJ...541..701G} Goldman, I.\ 2000, \apj, 541, 
701 

\bibitem[Kerr, Hindman, \& Robinson(1954)]{kerr54} Kerr, F. J.,
Hindman, J. F., Robinson, B. J.\ 1954, Aust. Jp. 7, 297

\bibitem[Kobulnicky \& Dickey(1999)]{1999AJ....117..908K} Kobulnicky, H. A.,
  Dickey, J. M.,\ 1999, \aj, 117, 908

\bibitem[Lazarian \& Esquivel(2003)]{2003ApJ...592L..37L} Lazarian, A.~\& 
Esquivel, A.\ 2003, \apjl, 592, L37 

\bibitem[Lazarian \& Pogosyan(2000)]{2000ApJ...537..720L} Lazarian, A.~\& 
Pogosyan, D.\ 2000, \apj, 537, 720 

\bibitem[Mathewson, Schwarz, \& Murray(1977)]{1977ApJ...217L...5M} 
Mathewson, D.~S., Schwarz, M.~P., \& Murray, J.~D.\ 1977, \apjl, 217, L5 

\bibitem[McGee \& Newton]{1986PASAu...6..471M} McGee, R. X., Newton, L. M.,\
  1986, PASAu, 6,471

\bibitem[Minter(2002)]{2002stdd.conf..221M} Minter, A.\ 2002, ASP 
Conf.~Ser.~276: Seeing Through the Dust: The Detection of HI and the 
Exploration of the ISM in Galaxies, 221 

\bibitem[Miville-Desch{\^ e}nes, Levrier, \& 
Falgarone(2003)]{2003ApJ...593..831M} Miville-Desch{\^ e}nes, M.-A., 
Levrier, F., \& Falgarone, E.\ 2003, \apj, 593, 831 

\bibitem[Muller, Staveley-Smith, Zealey, \& Stanimirovi{\'
c}(2003)]{2003MNRAS.339..105M} Muller, E., Staveley-Smith, L., Zealey,
W., \& Stanimirovi{\' c}, S.\ 2003, \mnras, 339, 105

\bibitem[Padoan, Goodman, \& Juvela(2003)]{scf_slope} Padoan, P.,
Goodman, A.~A., \& Juvela, M.\ 2003, \apj, 588, 881

\bibitem[Padoan et al.(2001)]{scf_lmc} Padoan, P., Kim, S., Goodman,
A., \& Staveley-Smith, L.\ 2001, \apjl, 555, L33

\bibitem[Padoan, Rosolowsky, \& Goodman(2001)]{scf2} Padoan, P.,
Rosolowsky, E.~W., \& Goodman, A.~A.\ 2001, \apj, 547, 862 \

\bibitem[Rosolowsky et al.(1999)]{scf1} Rosolowsky, E.~W., Goodman,
A.~A., Wilner, D.~J., \& Williams, J.~P.\ 1999, \apj, 524, 887

\bibitem[{Ru\v{z}i\v{c}ka}(2003)]{ruzicka03} Ru\v{z}i\v{c}ka, A.,
2003, A\&SS, 484, 519

\bibitem[Scalo \& Elmegreen(2004)]{astro-ph/0404452}Scalo, J., Elmegreen,
  B. G., 2004 astro-ph/0404452

\bibitem[Stanimirovi{\' c}, Dickey, Kr{\v c}o, \& 
Brooks(2002)]{2002ApJ...576..773S} Stanimirovi{\' c}, S., Dickey, J.~M., 
Kr{\v c}o, M., \& Brooks, A.~M.\ 2002, \apj, 576, 773 

\bibitem[Stanimirovi{\' c} \& Lazarian(2001)]{2001ApJ...551L..53S} 
Stanimirovi{\' c}, S.~\& Lazarian, A.\ 2001, \apjl, 551, L53 

\bibitem[Stanimirovic et al.(1999a)]{1999MNRAS.302..417S} Stanimirovic, S., 
Staveley-Smith, L., Dickey, J.~M., Sault, R.~J., \& Snowden, S.~L.\ 1999a, 
\mnras, 302, 417 

\bibitem[Stanimirovic et al.(1999b)]{1999IAUS..190..103S} Stanimirovic, S., 
Staveley-Smith, L., Sault, R.~J., Dickey, J.~M., \& Snowden, S.~L.\ 1999a, 
IAU Symp.~190: New Views of the Magellanic Clouds, 190, 103 

\bibitem[Stanimirovi{\' c} et~al.(2003)]{stan03} {Stanimirovi{\' c}},
S., {Weisberg}, J., {Dickey}, J.~M., {de la Fuente}, A., {Devine}, K.,
{Hedden}, A., \& {Anderson}, S.~B. 2003, \apj, 592, 953

\bibitem[Stanimirovi{\' c} (1999)]{1251} Stanimirovi{\' c}, S, PHDT, University
of Western Sydney, 1999

\bibitem[Stavely-Smith, Kim, Putman, \& Stanimirovi{\' 
c}(1998)]{1998RvMA...11..117S} Stavely-Smith, L., Kim, S., Putman, M., \& 
Stanimirovi{\' c}, S.\ 1998, Reviews of Modern Astronomy, 11, 117 
\end{thebibliography}
\end{document}